# Qubit Field Theory

## David Deutsch

Centre for Quantum Computation,
The Clarendon Laboratory
University of Oxford, Oxford OX1 3PU, UK

and

Centre for Quantum Computation,
Department of Applied Mathematics and Theoretical Physics
University of Cambridge, Cambridge CB3 0WA, UK

*The canonical commutation relations of quantum field theory require all pairs of observables located in spacelike-separated regions to commute. In the theory as it is currently constituted, this implies that the information-carrying capacity of a finite volume of space is infinite. Yet Bekenstein's bound gives us strong reason to believe that it is finite. A class of quantum field theories is presented in which observables localised in spacelike-separated regions do not necessarily commute, but which nevertheless has no physical pathologies.*

## 1 Motivation

Is there an upper bound on the amount of information that can be stored in a given finite region $\mathsf{R}$ of space? A naive application of quantum field theory would imply that there is not: for instance, the observables $\hat{\varphi}(\mathbf{x},t)$ of a scalar quantum field commute at spacelike separations,

$$\left[\hat{\varphi}(\mathbf{x},t),\hat{\varphi}(\mathbf{x}',t)\right] = 0, \tag{1}$$

where $\mathbf{x}$ and $\mathbf{x}'$ denote position on a spacelike hypersurface $t$ of a spacetime $\mathsf{M}$. Therefore in any region $\mathsf{R}$ on such a hypersurface one can find an arbitrarily large number of mutually commuting observables, built out of field operators from mutually disjoint sub-regions of R (Fig. 1). One can choose any two distinct eigenvalues of such an observable in each sub-region to represent 0 and 1, and hence it would seem that one could simultaneously prepare all those observables of R with arbitrary values in {0,1} and later measure them again with arbitrary accuracy. Hence, for an instant, one would have stored an arbitrarily large number of bits of information in R.

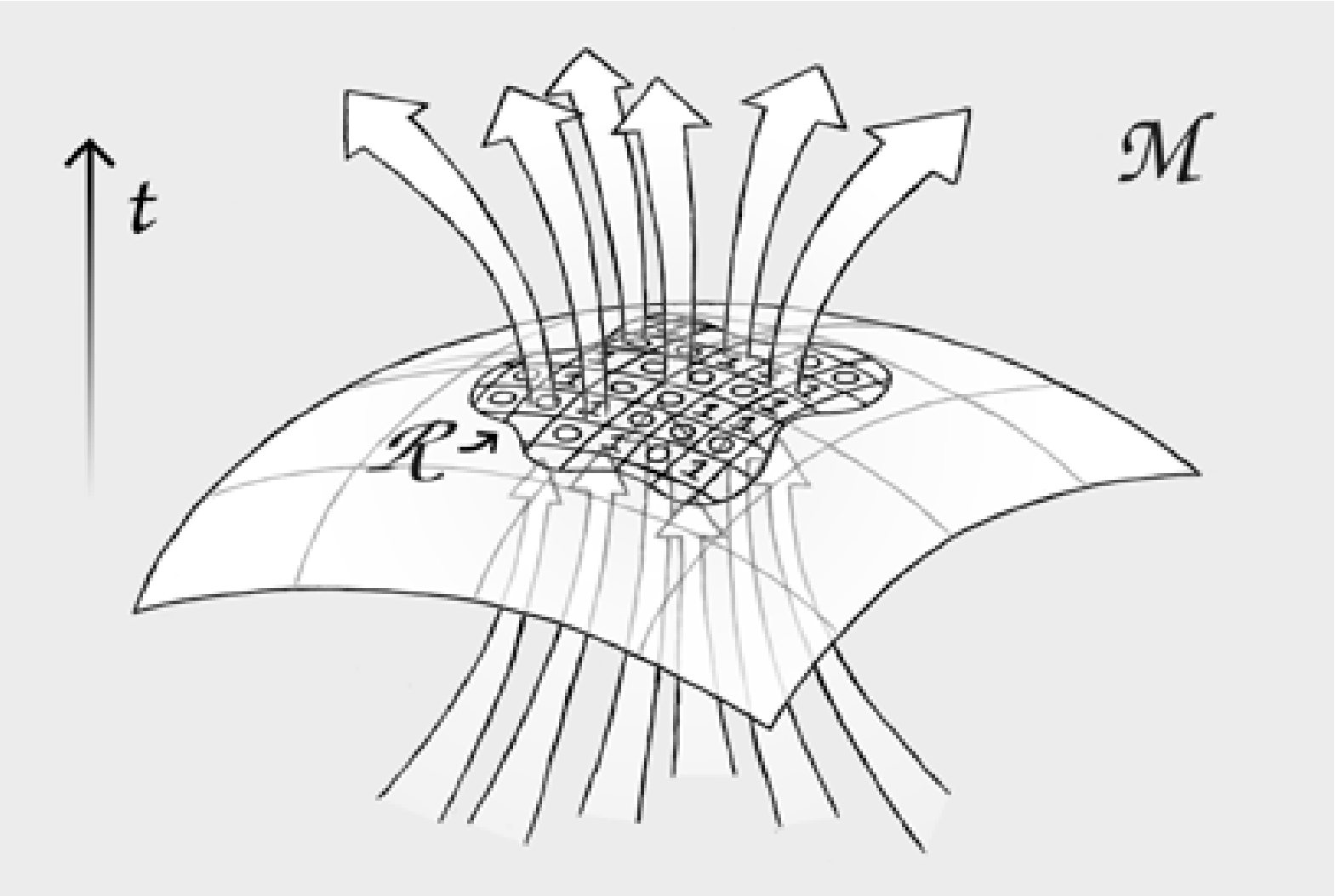

Fig. 1: Storing information in a region R

However, although no counterexample has ever been observed to the theory that leads to this conclusion, there are good reasons to believe that it becomes inaccurate in precisely the regimes contemplated in the above construction – regimes of very dense information storage. Roughly speaking, states in which the field varies rapidly in space tend to be states of high energy, and when the energy becomes sufficiently high, the field in question must begin to interact significantly with the gravitational field. Eventually, and certainly before the energy exceeds $(A/\pi)^{\frac{1}{2}} c^4 / G$, where $A$ is the area of a surface enclosing R, $c$ is the speed of light and $G$ is Newton's constant, the region becomes a black hole and no information can be retrieved from it. The information storage capacity of a finite region is thought be bounded by Bekenstein's (1973) formula, as refined by Hawking (1975) for the entropy $S$ of any region enclosed by a surface of area $A$:

$$S \le \frac{kc^4 A}{4\hbar G}, \tag{2}$$

where $k$ is Boltzmann's constant and $\hbar$ is Planck's reduced constant. The equality holds if and only if the system is a static black hole, so (2) says that a black hole has at least as much entropy as any other object that could be enclosed in the same surface. The Hilbert space of the quantum field inside such a region therefore cannot have dimension higher than $e^{S/k}$, and so the region cannot hold more than $\frac{c^4 A}{4\hbar G \ln 2}$ bits. Bekenstein's arguments in favour of such a bound are based on intuitively compelling thought experiments that seem to be insensitive to the details of the underlying quantum theory of gravity.

Unfortunately, no known quantum field theory actually displays this behaviour. So let us suppose that (2) is true. Is it possible to obtain a viable quantum field theory that satisfies it by relaxing the condition that observables constructed from field quantities at spacelike separations commute? That is the question that I shall explore in this paper.

My approach will be to make the least possible change to existing quantum field theory, subject to the above relaxation.

## 2 Qubit field theories

Consider a field of identically-constituted quantum physical systems on spacetime. That is to say, each event is associated with one such system, and each of the systems has the same algebra of observables and obeys the same dynamical law. The field is continuous in the sense that the corresponding fields of physical quantities (such as identically-named observables) of each system are all continuous, and also differentiable as many times as we need. Conventional quantum fields do not quite fit this simple description – they are distribution-valued and do not have observables, let alone physical systems, located at individual events – but the theory I shall present here does.

This simplest possible physical system is one whose observables have the algebra of $2\times 2$ Hermitian matrices. Such systems are known as *qubits*. And so the simplest quantum field of the type that I am seeking is a field of qubits. Let $\mathsf{Q}_x$ be the qubit at event $x$. $\mathsf{Q}_x$ is conveniently described in the Heisenberg Picture (see Deutsch and Hayden (2000)) in terms of its three representative observables $\hat{q}_j(x)$ $(1 \le j \le 3)$ which obey the Pauli algebra

$$\hat{q}_j(x)\hat{q}_k(x) = \delta_{jk}\hat{1} + i\varepsilon_{jk}{}^{l}\hat{q}_l(x). \tag{3}$$

Here and throughout I use the Einstein summation convention in which an index occurring once in the superscript and once in the subscript position in a product is summed over all its possible values. The index $l$ in (3) has been raised merely to indicate that it is to be summed over in accordance with that convention.

All observables of $\mathsf{Q}_x$ are linear combinations of the $\hat{q}_j(x)$ and the unit observable $\hat{1}$, with real coefficients. In a conventional quantum field theory one would now impose some commutation relation including the condition that $[\hat{q}_j(x), \hat{q}_k(x')] = 0$ whenever $x$ and $x'$ are separated by a spacelike interval. Here, I shall not, a priori, impose any commutation relation on observables at different events. (Readers who are concerned that this might permit superluminal signalling should note that the main result of Deutsch and Hayden (2000) – that there are no non-local effects in quantum physics – applies equally to the theory to be described here, provided that the dynamical evolution of the fields is local (see below).)

Thus there is a 2-dimensional Hilbert space $\boldsymbol{H}_x$ associated with each event in spacetime: Choose a particular event $x$. The observables of the qubit $\mathsf{Q}_x$, in isolation, can be thought of as Hermitian operators on $\boldsymbol{H}_x$ and as such their algebra (3) can be represented in terms of 2-dimensional matrices, such as the Pauli matrices

$$\begin{gathered}\sigma_1=\begin{pmatrix}0 & 1\\ 1 & 0\end{pmatrix},\ \sigma_1=\begin{pmatrix}0 & -i\\ i & 0\end{pmatrix},\ \sigma_1=\begin{pmatrix}1 & 0\\ 0 & -1\end{pmatrix},\\ \hat{q}_j(x)\leftrightarrow\sigma_j,\end{gathered} \tag{4}$$

where the symbol "$\leftrightarrow$" denotes "represents or is represented by". Now, $\boldsymbol{H}_x$ is a subspace of the overall Hilbert space $\boldsymbol{H}$ of the field. However, because of the non-commutativity, $\boldsymbol{H}$ is not a continuously infinite product of such subspaces, or anything of the sort. In fact we are expecting the field in any finite region of spacetime to have a finite-dimensional matrix representation.

For each event $x$, the set of all observables that commute with every observable of $\mathsf{Q}_x$ form a vector space $\overline{\mathbf{O}}_x$ which is a subspace of the vector space $\mathbf{O}$ of all observables of the field. The projector $\vec{\Pi}_x$ into the space $\overline{\mathbf{O}}_x$ is defined by its effect on an arbitrary observable $\hat{A}$ as follows:

$$\vec{\Pi}_x\hat{A}\overset{\text{def}}{=}\tfrac{1}{4}\left(\hat{A}+\hat{q}_j(x)\hat{A}\hat{q}^j(x)\right). \tag{5}$$

The arrow in $\overrightarrow{\Pi}_x$ indicates that it is a super-operator, *i.e.* a linear operator mapping $\mathbf{O}$ to itself. Likewise $\overrightarrow{1}-\overrightarrow{\Pi}_x$, where $\overrightarrow{1}$ is the identity super-operator, is the projector into the space $\mathbf{O}_x$ of observables orthogonal to all those in $\overline{\mathbf{O}}_x$. Thus an arbitrary observable $\hat{A}$ can be written in terms of its projections into $\overline{\mathbf{O}}_x$ and $\mathbf{O}_x$ via the identity

$$\hat{A}\equiv\overrightarrow{\Pi}_x\hat{A}+\left(\overrightarrow{1}-\overrightarrow{\Pi}_x\right)\hat{A}. \tag{6}$$

The algebra of the observables of $\overline{\mathbf{O}}_x$, taken in isolation, have a representation in terms of $\frac{1}{2}N$-dimensional matrices where $N$ is the dimension of $\boldsymbol{H}$, so they can be thought of as operators on the $\frac{1}{2}N$-dimensional quotient Hilbert space $\boldsymbol{H}/\boldsymbol{H}_x$. Thus, at each event $x$, the local qubit $\mathsf{Q}_x$ defines a product structure on $\boldsymbol{H}$, partitioning it into 2- and $\frac{1}{2}N$-dimensional factors $\boldsymbol{H}_x$ and $\boldsymbol{H}/\boldsymbol{H}_x$. Incidentally, this shows that $N$ must be even. In terms of this product structure we can represent the identity (6) as

$$\hat{A}\leftrightarrow \mathrm{I}\otimes A_0+\sigma_j\otimes A_j, \tag{7}$$

where I is the 2-dimensional unit matrix and $A_0$ and the $A_j$ are $\frac{1}{2}N$-dimensional matrices, and

$$\begin{aligned}\mathrm{I}\otimes A_0&\leftrightarrow\overrightarrow{\Pi}_x\hat{A}\\ \mathrm{I}\otimes A_j&\leftrightarrow\tfrac{1}{4}\left(\left\{\hat{A},\hat{q}_j(x)\right\}+i\varepsilon_j{}^{kl}\hat{q}_k(x)\hat{A}\hat{q}_l(x)\right).\end{aligned} \tag{8}$$

Now, since the qubits all have the same algebra of observables, the correspondingly-named observables of different qubits must be related by unitary transformations as follows:

$$\hat{q}_j(x)=U^{\dagger}(x)\hat{q}_j(0)U(x), \tag{9}$$

$$U^{\dagger}(x)U(x)=\hat{1}, \tag{10}$$

where ‘0’ represents some event arbitrarily chosen as an origin of coordinates. Hence

$$\hat{q}_{j;\mu}(x)=i[\hat{J}_{\mu}(x),\hat{q}_{j}(x)], \tag{11}$$

where the semicolon denotes covariant differentiation and

$$\hat{J}_{\mu}(x)\overset{\text{def}}{=}-iU^{\dagger}_{;\mu}(x)U(x). \tag{12}$$

$\hat{J}_{\mu}(x)$ is Hermitian because of (10). In view of (11), it acts as a sort of Hamiltonian in this theory, but with two important differences from conventional Hamiltonians: first, it is a field on spacetime, while conventional Hamiltonians are global quantities associated with each spacelike hypersurface; and second, it is a spacetime *vector* field while the conventional Hamiltonian is a scalar. Differentiating (12) again and rearranging, we obtain a commutation relation that this Hamiltonian must obey:

$$[\hat{J}_{\mu}(x),\hat{J}_{\nu}(x)]=i\left(\hat{J}_{\mu;\nu}(x)-\hat{J}_{\nu;\mu}(x)\right). \tag{13}$$

$\hat{J}_{\mu}$ is one of a family of Hamiltonians that would all generate identical motions (11). The difference between any two such Hamiltonians at any event $x$ must commute with all three of the $\hat{q}_{j}(x)$ and hence lie in $\overline{\mathbf{O}}_{x}$. In fact one such Hamiltonian is

$$\hat{H}_{\mu}(x)\overset{\text{def}}{=}\left(\vec{1}-\vec{\Pi}_{x}\right)\hat{J}_{\mu}(x), \tag{14}$$

for it follows from (14), (11) and (3) that

$$\hat{H}_{\mu}(x)=-\tfrac{i}{4}\hat{q}_{j;\mu}(x)\hat{q}^{j}(x), \tag{15}$$

and that

$$\hat{q}_{j;\mu}(x) = i[\hat{H}_\mu(x), \hat{q}_j(x)]. \tag{16}$$

Now, (15) and (16) do not specify the dynamics of the field: on the contrary, the above derivation applies no matter what the dynamics are, so long as they are unitary. To specify the dynamics, I shall now seek an equation of motion for such a field – specifically for a free field, in the sense that it does not interact with other fields and is not affected by any external potential.

The criteria for an acceptable equation of motion for such a field presumably include:

- that it be generally covariant.
- that it give rise to unitary evolution in the sense (9);
- that it have a well-posed initial-value problem;
- that it be local – *i.e.* that it refer only to $\hat{q}_j(x)$ and a finite number of its spacetime derivatives at one event.

I shall consider only the case where there is no preferred direction in the qubit's internal 3-space (corresponding to the indices *j*). Since the equation is going to be solved for $\hat{q}_j(x)$, given suitable initial data, then by a crude count of degrees of freedom, we can expect it to set to zero a sum of terms each of which is, like $\hat{q}_j(x)$, a spacetime scalar with a single index *j*. On this assumption, all other internal qubit- and spacetime indices occurring in the equation must be contracted, and there can be no term independent of the $\hat{q}_j(x)$ and its derivatives because no such term can have the requisite index. Because of (3), any product of the form $\hat{q}_j(x)\hat{q}_k(x)\ldots$ can be replaced by a sum of terms each containing at most a single $\hat{q}_j(x)$ at that location in the product. In particular, therefore, the only term not containing derivatives in the equation must be a multiple of $\hat{q}_j(x)$ itself.

Using the derivative of (3), we can reduce any product of fields and their first derivatives to a sum of terms in which the derivatives are all consecutive. There must be an even number of such derivatives in any term of the sum, because otherwise their spacetime indices could not all be contracted. Similarly, to allow all but one of the qubit indices to be contracted, there must be an odd number of factors carrying such an index. From (3), we can always replace an expression containing such a contracted pair of first derivatives by one containing no first derivatives:

$$\hat{q}_j^{;\mu}(x)\hat{q}_{k;\mu}(x) = \tfrac{1}{2}\Big(i\varepsilon_{jk}{}^{l}\,\Box\hat{q}_l(x) - \Box\hat{q}_j(x)\hat{q}_k(x) - \hat{q}_j(x)\Box\hat{q}_k(x)\Big). \tag{17}$$

Similarly, we can always replace terms containing four or more first-derivative factors by terms containing only second derivatives. Let us therefore deal with terms of this type under the heading of second-order differential equations.

The simplest of these are equations of the first degree in the second derivative. In such equations the second derivatives appear only as linear combinations, with real constant coefficients, of the six terms:

$$\left.\begin{array}{c}
\Box\hat{q}_j(x) \\
\hat{q}_k(x)\Box\hat{q}_j(x)\hat{q}^k(x) \\
\hat{q}^k(x)\Box\hat{q}_k(x)\hat{q}_j(x) + \hat{q}_j(x)\Box\hat{q}_k(x)\hat{q}^k(x) \\
i\Big(\hat{q}^k(x)\Box\hat{q}_k(x)\hat{q}_j(x) - \hat{q}_j(x)\Box\hat{q}_k(x)\hat{q}^k(x)\Big) \\
\varepsilon_j{}^{kl}\Big(\hat{q}_k(x)\Box\hat{q}_l(x) + \Box\hat{q}_l(x)\hat{q}_k(x)\Big) \\
i\varepsilon_j{}^{kl}\Big(\hat{q}_k(x)\Box\hat{q}_l(x) - \Box\hat{q}_l(x)\hat{q}_k(x)\Big)
\end{array}\right\}. \tag{18}$$

Thus, in terms of the following super-operators:

$$\left.\begin{aligned}
&\vec{\Omega}_x^{(0)}\hat{A}_j \overset{\text{def}}{=} \hat{A}_j && \left(\text{i.e. } \vec{\Omega}_x^{(0)} = \vec{1}\right)\\
&\vec{\Omega}_x^{(1)}\hat{A}_j \overset{\text{def}}{=} \hat{q}_k(x)\hat{A}_j\hat{q}^k(x) && \left(\text{i.e. } \vec{\Omega}_x^{(1)} = 4\vec{\Pi}_x - \vec{1}\right)\\
&\vec{\Omega}_x^{(2)}\hat{A}_j \overset{\text{def}}{=} \varepsilon_j{}^{kl}\left(\hat{q}_k(x)\hat{A}_l + \hat{A}_l\hat{q}_k(x)\right)\\
&\vec{\Omega}_x^{(3)}\hat{A}_j \overset{\text{def}}{=} i\varepsilon_j{}^{kl}\left(\hat{q}_k(x)\hat{A}_l - \hat{A}_l\hat{q}_k(x)\right)\\
&\vec{\Omega}_x^{(4)}\hat{A}_j \overset{\text{def}}{=} \hat{q}^k(x)\hat{A}_k\hat{q}_j(x) + \hat{q}_j(x)\hat{A}_k\hat{q}^k(x)\\
&\vec{\Omega}_x^{(5)}\hat{A}_j \overset{\text{def}}{=} i\left(\hat{q}^k(x)\hat{A}_k\hat{q}_j(x) - \hat{q}_j(x)\hat{A}_k\hat{q}^k(x)\right)
\end{aligned}\right\} \tag{19}$$

the most general equation of motion of the type we are considering is:

$$\lambda_\alpha\vec{\Omega}_x^{(\alpha)}\Box q_j(x) + \mu\hat{q}_j(x) = 0, \tag{20}$$

where the index $\alpha$ ranges from 0 to 5 and the $\lambda_\alpha$ and $\mu$ are real constants. Note that

$$\left.\begin{aligned}
&\vec{\Omega}_x^{(0)}q_j(x) = q_j(x)\\
&\vec{\Omega}_x^{(1)}q_j(x) = -q_j(x)\\
&\vec{\Omega}_x^{(2)}q_j(x) = 0\\
&\vec{\Omega}_x^{(3)}q_j(x) = -4q_j(x)\\
&\vec{\Omega}_x^{(4)}q_j(x) = 6q_j(x)\\
&\vec{\Omega}_x^{(5)}q_j(x) = 0
\end{aligned}\right\}, \tag{21}$$

Now, the super-operators $\vec{\Omega}_x^{(\alpha)}$ have the following composition table:

| $\vec{\Omega}_x^{(\alpha)}\vec{\Omega}_x^{(\beta)}$ | | $\vec{\Omega}_x^{(\alpha)}$ | | | | | |
|---|---|---|---|---|---|---|---|
| | $\beta \backslash \alpha$ | 0 | 1 | 2 | 3 | 4 | 5 |
| $\vec{\Omega}_x^{(\beta)}$ | 0 | $\vec{\Omega}_x^{(0)}$ | $\vec{\Omega}_x^{(1)}$ | $\vec{\Omega}_x^{(2)}$ | $\vec{\Omega}_x^{(3)}$ | $\vec{\Omega}_x^{(4)}$ | $\vec{\Omega}_x^{(5)}$ |
| | 1 | $\vec{\Omega}_x^{(1)}$ | $3\vec{\Omega}_x^{(0)}+2\vec{\Omega}_x^{(1)}$ | $\vec{\Omega}_x^{(2)}-2\vec{\Omega}_x^{(5)}$ | $-\vec{\Omega}_x^{(3)}$ | $2(\vec{\Omega}_x^{(0)}+\vec{\Omega}_x^{(1)})-\vec{\Omega}_x^{(4)}$ | $-2\vec{\Omega}_x^{(2)}+\vec{\Omega}_x^{(5)}$ |
| | 2 | $\vec{\Omega}_x^{(2)}$ | $\vec{\Omega}_x^{(2)}+2\vec{\Omega}_x^{(5)}$ | $-4\vec{\Omega}_x^{(0)}-2\vec{\Omega}_x^{(1)}+\vec{\Omega}_x^{(3)}+\vec{\Omega}_x^{(4)}$ | $-\vec{\Omega}_x^{(2)}+\vec{\Omega}_x^{(5)}$ | $-\vec{\Omega}_x^{(2)}+3\vec{\Omega}_x^{(5)}$ | $-2\vec{\Omega}_x^{(1)}-\vec{\Omega}_x^{(3)}-\vec{\Omega}_x^{(4)}$ |
| | 3 | $\vec{\Omega}_x^{(3)}$ | $-\vec{\Omega}_x^{(3)}$ | $-\vec{\Omega}_x^{(2)}-\vec{\Omega}_x^{(5)}$ | $4\vec{\Omega}_x^{(0)}-2\vec{\Omega}_x^{(1)}-\vec{\Omega}_x^{(3)}+\vec{\Omega}_x^{(4)}$ | $2\vec{\Omega}_x^{(1)}+\vec{\Omega}_x^{(3)}-3\vec{\Omega}_x^{(4)}$ | $-\vec{\Omega}_x^{(2)}-\vec{\Omega}_x^{(5)}$ |
| | 4 | $\vec{\Omega}_x^{(4)}$ | $2(\vec{\Omega}_x^{(0)}+\vec{\Omega}_x^{(1)})-\vec{\Omega}_x^{(4)}$ | $-\vec{\Omega}_x^{(2)}-3\vec{\Omega}_x^{(5)}$ | $2\vec{\Omega}_x^{(1)}+\vec{\Omega}_x^{(3)}-3\vec{\Omega}_x^{(4)}$ | $8\vec{\Omega}_x^{(0)}-2\vec{\Omega}_x^{(1)}-5\vec{\Omega}_x^{(3)}+\vec{\Omega}_x^{(4)}$ | $-3\vec{\Omega}_x^{(2)}-\vec{\Omega}_x^{(5)}$ |
| | 5 | $\vec{\Omega}_x^{(5)}$ | $2\vec{\Omega}_x^{(2)}+\vec{\Omega}_x^{(5)}$ | $2\vec{\Omega}_x^{(1)}+\vec{\Omega}_x^{(3)}+\vec{\Omega}_x^{(4)}$ | $\vec{\Omega}_x^{(2)}-\vec{\Omega}_x^{(5)}$ | $3\vec{\Omega}_x^{(2)}-\vec{\Omega}_x^{(5)}$ | $4\vec{\Omega}_x^{(0)}+2\vec{\Omega}_x^{(1)}-\vec{\Omega}_x^{(3)}-\vec{\Omega}_x^{(4)}$ |

Table 1: Composition of the super-operators $\vec{\Omega}_x^{(\alpha)}$

In other words,

$$\vec{\Omega}_x^{(\alpha)}\vec{\Omega}_x^{(\beta)} = c^{\alpha\beta}{}_{\gamma}\vec{\Omega}_x^{(\gamma)}, \tag{22}$$

where the 216 real constants $c^{\alpha\beta}{}_{\gamma}$ (actually we see that they are all integers) can be read off from Table 1. Since composition of the super-operators is associative, and $\vec{\Omega}_x^{(0)}$ is the unit element under composition, the set of all such super-operators (or correspondingly, of all differential operators of the form $\lambda_\alpha\vec{\Omega}_x^{(\alpha)}\Box$) constitutes a six-dimensional manifold with the structure of a Lie monoid (something with all the properties of a Lie group except that not every element has an inverse). Consider first the elements that do have inverses. These form a group. Given (19) and (21), any equation of the form (20) involving one such element can also be written in terms of any other. So, in particular, all such equations are equivalent to

$$(\Box + \mu)\hat{q}_j(x) = 0, \tag{23}$$

for some constant $\mu$. Let me call these 'equations of motion of type I'.

For the element $\lambda_\alpha\vec{\Omega}_x^{(\alpha)}$ to have an inverse $\bar{\lambda}_\alpha\vec{\Omega}_x^{(\alpha)}$, it is necessary and sufficient that

$$\lambda_\alpha\bar{\lambda}_\beta c^{\alpha\beta}{}_{\gamma} = \delta^0_\gamma, \tag{24}$$

which is equivalent to the requirement that the matrix $\lambda_\alpha c^{\alpha\beta}{}_\gamma$ be non-singular. Hence, for all equations of the form (20) that do *not* reduce to (23), the coefficients $\lambda_\alpha$ satisfy the homogeneous sixth degree polynomial equation

$$\det\left(\lambda_\alpha c^{\alpha\beta}{}_\gamma\right) = 0, \tag{25}$$

which is

$$(\lambda_0 - \lambda_1 + 2\lambda_3)(\lambda_0 - \lambda_1 - 4\lambda_3 + 6\lambda_4)\left(8\lambda_5^2 + 8\lambda_4^2 - 8\lambda_2^2 + 3\lambda_1^2 - \lambda_0^2 + 6\lambda_1\lambda_3 + 14\lambda_1\lambda_4 + 4\lambda_3\lambda_4 + 2\lambda_0(\lambda_4 + \lambda_3 - \lambda_1)\right)^2 = 0. \tag{26}$$

Let me call equations for which the first, second or third factor respectively in (26) vanishes, type II, III or IV equations of motion respectively. Equations for which the last two, the outer two, or the first two vanish are of types V, VI and VII, and equations for which all three factors vanish are of type VIII. In tentative imitation of conventional field theory, let me call the fields 'massless' if $\mu = 0$ and 'massive' if $\mu \neq 0$.

## 3 A model theory

We can classify solutions of equations of the form (20) according to the dimension of their smallest matrix representation. Clearly no 1-dimensional matrices can satisfy (3). If the $\hat{q}_j(x)$ are 2-dimensional, then $\vec{\Pi}_x = \vec{1}$ for all $x$ (see (5)), or, in terms of the $\vec{\Omega}_x^{(\alpha)}$:

$$\vec{\Omega}_x^{(1)} = 3\vec{\Omega}_x^{(0)}. \tag{27}$$

Hence, from the first two equations of (21), there are no 2-dimensional solutions either. Since the dimension must be even, it must be at least 4.

The space of all $4\times 4$ Hermitian matrices is spanned by the sixteen matrices $\sigma_j\otimes\sigma_k$, $I\otimes\sigma_k$, $\sigma_j\otimes I$ and $I\otimes I$ where the $\sigma_j$ are the Pauli matrices and $I$ is the $2\times 2$ unit matrix. The super-operator $\vec{W}$ defined by

$$\left.\begin{aligned} \vec{W}\hat{A} &\overset{\text{def}}{=} W^{\dagger}\hat{A}W \\ W &\overset{\text{def}}{=} \tfrac{1}{2}\left(I\otimes I+\sigma_j\otimes\sigma^j\right) \end{aligned}\right\} \tag{28}$$

is called the *swap* super-operator because $\vec{W}(A\otimes B)=B\otimes A$, where $A$ and $B$ are arbitrary $2\times 2$ Hermitian matrices. If $\phi$ is any real number, the $\phi$th power of $\vec{W}$ is given by

$$\left.\begin{aligned} \vec{W}^{\phi}\hat{A} &= W^{-\phi}\hat{A}W^{\phi} \\ W^{\phi} &= \tfrac{1}{4}\left(\left(3+e^{i\pi\phi}\right)I\otimes I+\left(1-e^{i\pi\phi}\right)\sigma_j\otimes\sigma^j\right) \end{aligned}\right\}. \tag{29}$$

I am going to seek a solution of (20) in the form

$$\hat{q}_j(x)=\vec{W}^{\phi(x)}\left(\sigma_j\otimes I\right), \tag{30}$$

where $\phi(x)$ is a real c-number field. We have, from (30) and (29):

$$\hat{q}_j(x)=\frac{1}{2}\left(\left(1-\cos\pi\phi(x)\right)I\otimes\sigma_j+\left(1+\cos\pi\phi(x)\right)\sigma_j\otimes I-\sin\pi\phi(x)\varepsilon_j{}^{kl}\sigma_k\otimes\sigma_l\right), \tag{31}$$

and hence

$$\begin{aligned} \Box\hat{q}_j(x)=&\frac{\pi}{2}\left(\sin\pi\phi(x)\left(I\otimes\sigma_j-\sigma_j\otimes I\right)-\cos\pi\phi(x)\varepsilon_j{}^{kl}\sigma_k\otimes\sigma_l\right)\Box\phi(x)+ \\ &+\frac{\pi^2}{2}\left(\cos\pi\phi(x)\left(I\otimes\sigma_j-\sigma_j\otimes I\right)+\sin\pi\phi(x)\varepsilon_j{}^{kl}\sigma_k\otimes\sigma_l\right)\phi_{,\mu}(x)\phi^{,\mu}(x). \end{aligned} \tag{32}$$

We see at once that the ansatz (30) cannot provide a non-trivial solution of a type I equation of motion. For if $(\Box+\mu)\hat{q}_j(x)$ is to vanish, comparing coefficients of the matrices in (31) and (32) shows that $\mu=0$, and that $\Box\phi(x)=0$ and $\phi_{;\mu}(x)\phi^{;\mu}(x)=0$. But in a general curved spacetime there are no non-constant scalar fields with that property.

However, if we choose any $\phi(x)$ obeying $\Box\phi(x)=0$, the ansatz gives

$$\Box\hat{q}_j(x)=\frac{\pi^2}{2}\Big(\cos\pi\phi(x)\big(I\otimes\sigma_j-\sigma_j\otimes I\big)+\sin\pi\phi(x)\epsilon_j{}^{kl}\sigma_k\otimes\sigma_l\Big)\phi_{;\mu}(x)\phi^{;\mu}(x), \qquad (33)$$

and hence

$$\left(\vec{\Omega}_x^{(2)}+\vec{\Omega}_x^{(5)}\right)\Box\hat{q}_j(x)=0, \qquad (34)$$

or, more explicitly,

$$\left\{\hat{q}^k(x),\left[\hat{q}_{[k}(x),\Box\hat{q}_{j]}(x)\right]\right\}=0. \qquad (35)$$

This is an equation of motion of type VIII (massless). Thus, for any real scalar field $\phi(x)$ with $\Box\phi(x)=0$, the ansatz (31) is a solution of any equation of that type. Using the composition table (Table 1) and acting on (34) with each super-operator in succession, we find that the most general such equation is

$$\left(\lambda_1\left(2\vec{1}-\vec{\Omega}_x^{(3)}-\vec{\Omega}_x^{(4)}\right)+\lambda_2\left(\vec{\Omega}_x^{(2)}+\vec{\Omega}_x^{(5)}\right)\right)\Box\hat{q}_j(x)=0. \qquad (36)$$

I shall leave the task of finding the general solution of (36), and of equations of types I-VII, as exercises for the reader.

## 4 Global and Local Quantities

As always in quantum theory, the expectation value of any observable $\hat{A}$ is $\langle\hat{A}\rangle = \mathrm{Tr}\hat{A}\hat{\rho}$, where $\hat{\rho}$ is a global q-number constant, the density operator. The variance $\langle\hat{A}^2\rangle - \langle\hat{A}\rangle^2$ vanishes if and only if $\hat{A}$ commutes with $\hat{\rho}$, in which case $\hat{A}$ is said to be *sharp*. In the conventional theory, the field is said to be in a *stationary state* if the Hamiltonian is sharp, for in that case no expectation values change with time (even though the observables themselves still do change in general). By analogy with this, we may define a qubit $\mathsf{Q}_x$ as being *instantaneously stationary* if there exists a timelike vector $n^\mu$ at $x$ such that

$$n^\mu \left\langle \left[ \hat{H}_\mu(x), \hat{q}_j(x) \right] \right\rangle = 0. \tag{37}$$

If a qubit is stationary everywhere on a worldline with tangent vector field $n^\mu(x)$, then it may be useful to identify the qubits on that worldline as being 'the same qubit over time'. Similarly if the expectation values of all the $\hat{q}_j(x)$ are constant throughout a spacelike region (including one of dimension lower than 3), or equivalently if (37) holds for all spacelike $n^\mu(x)$ in the region, then the field can be said to be *homogeneous* in that region.

We can also define a local density operator $\hat{\rho}(x)$ by taking the partial trace of $\hat{\rho}$ over the $\frac{1}{2}N$-dimensional subspace $\boldsymbol{H}/\boldsymbol{H}_x$ defined in section 2, as follows: given (7) and (8),

$$\begin{aligned} \hat{\rho}(x) &\overset{\text{def}}{\leftrightarrow} \left( \mathrm{Tr}_{\boldsymbol{H}/\boldsymbol{H}_x} \hat{\rho} \right) \otimes \mathrm{I}_\perp \\ &= \tfrac{1}{2}\left( \hat{1} + \langle \hat{q}_j(x) \rangle \hat{q}^j(x) \right) \end{aligned} \tag{38}$$

which has the property that if $\hat{A}(x)$ is any observable of $\mathsf{Q}_x$,

$$\langle \hat{A}(x) \rangle = \frac{\mathrm{Tr}\hat{A}(x)\hat{\rho}(x)}{\mathrm{Tr}\hat{\rho}(x)}. \tag{39}$$

(Thus $\hat{\rho}(x)$ as defined in (38) is an un-normalised density operator, and the normalisation factor $\mathrm{Tr}\hat{\rho}(x)$ is necessarily $\frac{1}{2}N$.) I write $\hat{\rho}(x)$ with a caret, as if it were an observable, but as in the conventional theory, it is equal to a different observable at each event, and does not generally evolve in a unitary way: for instance, its eigenvalues generally change as the qubit becomes more or less entangled with the rest of the field.

Now we encounter a major difference from conventional field theory: in the present theory, it is generically impossible for every qubit of the field simultaneously to be in an unentangled state (or, in particular, in a pure state), or even arbitrarily close to one; in fact, except where the field is stationary and homogeneous, only qubits at isolated events can be un-entangled. To prove this, note first that the condition for $\mathbf{Q}_x$ to be unentangled is that in the product representation (7) at $x$, $\hat{\rho} \leftrightarrow \rho(x) \otimes \rho_\perp$, where

$$\begin{aligned} \rho(x) \otimes \mathrm{I}_\perp &\leftrightarrow \hat{\rho}(x) \\ \mathrm{I} \otimes \rho_\perp &\leftrightarrow 2\vec{\Pi}_x \hat{\rho} \end{aligned} \tag{40}$$

In other words

$$\hat{\rho} = \tfrac{1}{2}\hat{\rho}(x)\vec{\Pi}_x \hat{\rho}, \tag{41}$$

or $\hat{D}(x) = 0$, where

$$\hat{D}(x) \stackrel{\mathrm{def}}{=} 3\hat{\rho} - \hat{q}_j(x)\hat{\rho}\hat{q}^j(x) - \left(\left\{\hat{q}_j(x), \hat{\rho}\right\} + i\varepsilon^{jkl}\hat{q}_l(x)\hat{\rho}\hat{q}_k(x)\right)\mathrm{Tr}\hat{\rho}\hat{q}^j(x). \tag{42}$$

If $\hat{D}(x)$ is to vanish along a line through $x$ whose tangent vector is $n^\mu$, then so must $n^\mu \hat{D}_\mu(x)$, and with it, each of the three coefficients $n^\mu \mathrm{Tr}\left(\hat{q}_m \hat{D}_\mu(x)\right)$. But when $\hat{D}(x)=0$,

$$\begin{aligned} \mathrm{Tr}\left(\hat{q}_m \hat{D}_\mu(x)\right) &= -4i\,\mathrm{Tr}\left(\hat{q}_m(x)\left[\hat{H}_\mu(x),\hat{\rho}\right]\right) \\ &= 4i\left\langle\left[\hat{H}_\mu(x),\hat{q}_m(x)\right]\right\rangle, \end{aligned} \tag{43}$$

and so the field has to be stationary and homogenous throughout any extended unentangled region.

Note that $\hat{\rho}(x)$ is the local *Heisenberg* density operator, not a Schrödinger density operator. In this theory, since there is no global Hamiltonian, there is no global Schrödinger Picture. That is to say, although for any particular qubit-over-time one can construct a Schrödinger Picture in the usual way, there is no way of reformulating the theory of the field as a whole in terms of observables that do not change with time and a global state that does. A simple reason for that is that in the present theory, it is possible for two observables that are equal at one time (*i.e.* have identical matrix representations at that time) to become unequal later.

Although there is no global Hamiltonian generating the dynamics, that does not preclude the existence of global conserved observables. For example, consider a qubit field of type I – *i.e.* one for which $(\square+\mu)\hat{q}_j(x)=0$. On an arbitrary spacelike hypersurface $\Sigma$, construct the observable

$$\hat{E}_\Sigma \overset{\mathrm{def}}{=} \int_\Sigma \hat{H}_\mu(x)\, d\Sigma^\mu . \tag{44}$$

From (15), we have

$$\begin{aligned}\hat{E}_\Sigma &= -\tfrac{1}{4} i \int_\Sigma \hat{q}_{j;\mu}(x)\hat{q}^j(x)\,d\Sigma^\mu \\ &= \tfrac{1}{8} i \int_\Sigma \hat{q}_j(x)\overleftrightarrow{\nabla}_\mu \hat{q}^j(x)\,d\Sigma^\mu\end{aligned} \quad . \tag{45}$$

and hence in the usual way, applying Green's theorem to the spacetime region between two such hypersurfaces, we find that $\hat{E}_\Sigma$ is hypersurface-independent – *i.e.* it is a conserved quantity.

Since a solution of an equation of type I is also a solution of equations of all the other types, it seems likely that at least some classes of solution of each type of equation exhibit global conserved quantities.

For general solutions, equations of motion of type VII also give rise to conserved quantities analogous to the above, but none of the other types do. That is because the possible integrands that are perfect divergences of a quantity containing one derivative and no free internal index are linear combinations of:

$$\begin{aligned}\left(\hat{q}_{j;\mu}(x)\hat{q}^j(x)\right)^{;\mu} &= \tfrac{1}{2}\left(\Box\hat{q}_j(x)\hat{q}^j(x) - \hat{q}^j(x)\Box\hat{q}_j(x)\right) \\ \text{and } \left(\varepsilon^{jkl}\hat{q}_j(x)\hat{q}_{k;\mu}(x)\hat{q}_l(x)\right)^{;\mu} &= 0,\end{aligned} \tag{46}$$

which gives nothing new. The ones with one free internal index are linear combinations of the $\left(\vec{\Omega}_x^{(\alpha)}\hat{q}_{j;\mu}(x)\right)^{;\mu}$:

$$\left.\begin{aligned}
\left(\vec{\Omega}_x^{(0)}\hat{q}_{j,\mu}(x)\right)^{,\mu} &= \square\hat{q}_j(x) \\
\left(\vec{\Omega}_x^{(1)}\hat{q}_{j,\mu}(x)\right)^{,\mu} &= -\square\hat{q}_j(x) \\
\left(\vec{\Omega}_x^{(2)}\hat{q}_{j,\mu}(x)\right)^{,\mu} &= \vec{\Omega}_x^{(2)}\square\hat{q}_j(x) \\
\left(\vec{\Omega}_x^{(3)}\hat{q}_{j,\mu}(x)\right)^{,\mu} &= -2\square\hat{q}_j(x) \\
\left(\vec{\Omega}_x^{(4)}\hat{q}_{j,\mu}(x)\right)^{,\mu} &= -4\square\hat{q}_j(x) \\
\left(\vec{\Omega}_x^{(5)}\hat{q}_{j,\mu}(x)\right)^{,\mu} &= \vec{\Omega}_x^{(2)}\square\hat{q}_j(x)
\end{aligned}\right\}, \tag{47}$$

from which it follows that $\int_\Sigma \vec{\Omega}_x^{(2)}\hat{q}_{j,\mu}(x)d\Sigma^\mu$, which equals $\int_\Sigma \vec{\Omega}_x^{(5)}\hat{q}_{j,\mu}(x)d\Sigma^\mu$, is a global conserved quantity for fields of type VII. The perfect divergences containing one derivative and two or more free internal indices again give nothing new.

Another global quantity that plays an important role in conventional quantum field theory is the action functional (which is a property of the whole field configuration in spacetime) and its associated Lagrangian (which is a property of the field configuration on hypersurfaces). Conventional theory is intentionally ambiguous in regard to whether the action is a c-number or a q-number (a feature that I have criticised in Deutsch (1984)), but in the present theory, which cannot usefully be regarded as the 'quantised' version of any classical field theory, we have to be specific. Consider a q-number Lagrangian density $\hat{\mathcal{L}}\left[\hat{q}_j(x),\hat{q}_{j,\mu}(x),\ldots\right]$ (in practice involving no higher derivatives than the first), and thence a q-number action functional of the form

$$\hat{S}\left[\hat{q}_j(x)\right] = \int_{\mathcal{M}} \hat{\mathcal{L}}\, dx, \tag{48}$$

where $dx$ denotes the covariant 4-volume element. From that, we can define a c-number action functional $S\left[\hat{q}_j(x)\right] = \mathrm{Tr}\hat{S}\left[\hat{q}_j(x)\right]$. In the q-number case, we require the action to be at an extremum with respect to variations of the form $\delta\hat{q}_j(x)=\delta q_j(x)\hat{1}$ where the $\delta q_j(x)$ are suitable infinitesimal c-number functions – even though such a variation will not preserve the commutation relations (3). In the case of a c-number action, we allow infinitesimal but otherwise arbitrary q-number variations $\delta\hat{q}_j(x)$, again not constrained by the commutation relations (3). A third possibility is that advocated in Deutsch (1984), of using a q-number action and q-number variations of the form $\delta\hat{q}_j(x)=i\varepsilon_j{}^{kl}\delta q_k(x)\hat{q}_l(x)$ which preserve the commutation relations. In all three cases we have to impose the commutation relations as a supplementary condition independent of the variational principle.

As in Section 2, let me restrict attention to theories that are generally covariant and have no preferred direction in the internal space. Although $\hat{q}_j(x)\hat{q}^j(x)$ and $\varepsilon^{jkl}\hat{q}_j(x)\hat{q}_k(x)\hat{q}_l(x)$ are both constants ($3\hat{1}$ and $6i\hat{1}$ respectively), that is a consequence of the commutation relations, and therefore those terms are available for Lagrangians for the first two action principles, where they give rise to terms $\mu\hat{q}_j(x)$ in the equation of motion. But for the Deutsch (1984) action principle, the qubit field has to be massless.

The simplest first-derivative Lagrangian density is $\hat{q}_{j,\mu}(x)\hat{q}^{j,\mu}(x)$. With this we can construct a q-number action

$$\hat{S}\left[\hat{q}_j(x)\right]=\int_{M}\hat{q}_{j,\mu}(x)\hat{q}^{j,\mu}(x)dx\,, \tag{49}$$

and the resulting equation of motion is simply $\Box\hat{q}_j(x)=0$, describing a qubit field of type I. We obtain the same equation of motion from the associated c-number action. The Deutsch (1984) action principle gives $\vec{\Omega}_x^{(2)}\Box\hat{q}_j(x)=0$, which is of type VII.

| Lagrangian Density | Action Principle | | |
|---|---|---|---|
| | c-number action, q-number variation | q-number action, c-number variation | q-number action, q-number variation |
| $\hat{q}_{j\mu}(x)\hat{q}^{j\mu}(x)$ | $\square\hat{q}_j(x)$<br>Type I | $\square\hat{q}_j(x)$<br>Type I | $\vec{\Omega}_x^{(2)}\square\hat{q}_j(x)$<br>Type VII |
| $\hat{q}_k(x)\hat{q}_{j\mu}(x)\hat{q}^{j\mu}(x)\hat{q}^k(x)$ | $\left(14\vec{\Omega}_x^{(0)}-\vec{\Omega}_x^{(3)}+\vec{\Omega}_x^{(4)}\right)\square\hat{q}_j(x)$<br>Type I | $\left(2\vec{\Omega}_x^{(0)}-\vec{\Omega}_x^{(3)}-\vec{\Omega}_x^{(4)}\right)\square\hat{q}_j(x)$<br>Type VIII | $\vec{\Omega}_x^{(2)}\square\hat{q}_j(x)$<br>Type VII |
| Linear combination of the above | $\left(\lambda\vec{\Omega}_x^{(0)}-\vec{\Omega}_x^{(3)}+\vec{\Omega}_x^{(4)}\right)\square\hat{q}_j(x)$<br>Type III ($\lambda$ = -10), IV ($\lambda$ = -2), VI ($\lambda$ = 2), or I (otherwise). | $\left(\lambda\vec{\Omega}_x^{(0)}-\vec{\Omega}_x^{(3)}-\vec{\Omega}_x^{(4)}\right)\square\hat{q}_j(x)$<br>Type IV ($\lambda$ = -6), VIII ($\lambda$ = 2), or I (otherwise). | As above. |

Table 2: Terms in the equations of motion generated by various Lagrangians and action principles

The next-simplest Lagrangian density is $\hat{q}_k(x)\hat{q}_{j\mu}(x)\hat{q}^{j\mu}(x)\hat{q}^k(x)$. This gives rise to equation-of-motion terms as shown in Table 2. Linear combinations of these two Lagrangian densities give rise, in the case of the first two action principles, to a one-parameter family of equations of motion whose types depend on the parameter, as is also summarised in Table 2.

## 5 Open questions

All such theories satisfy the criteria suggested in Section 2: they are generally covariant, local and unitary in the sense (9), and have well-posed initial-value problems. Furthermore, all the representations of any such field, of a given dimension, are unitarily related, and they are manifestly finite. All this makes them extremely well-behaved by the standards of quantum field theory. Further quantum field theories with these properties can evidently be constructed by considering fields with higher-dimensional local Hilbert spaces. Any such field can also be regarded as a finite set of interacting qubit fields.

That such theories should exist seems interesting from a theoretical point of view independently of whether they correspond to anything in nature. But if any of them do, then the whole physical world must consist of interacting qubit fields. For just as it is not possible to couple a classical dynamical system consistently to a quantum one, so it is not possible to couple a field that commutes at spacelike separations to one that does not. And the reason is fundamentally the same: Suppose that at time $t_1$ a pair of spacelike-separated but non-commuting Boolean observables $\hat{A}(t_1)$ and $\hat{B}(t_1)$ of a qubit field (with eigenvalues ±1) are measured, and that by a later time $t_2$ the outcomes are stored in a pair of Boolean observables $\hat{X}$ and $\hat{Y}$ of a conventional field. So $\hat{X}(t_1)$ and $\hat{Y}(t_1)$ commute, but even if the two measurement processes happen entirely within the respective spacelike-separated regions, $\hat{X}(t_2)$ must now be a function of $\hat{A}(t_1)$ (say $\hat{X}(t_2)=\hat{A}(t_1)\hat{X}(t_1)$) and must likewise be a function of $\hat{B}(t_1)$. So they can no longer commute, contradicting the supposition that the second field is a conventional one.

If qubit fields are realised in nature then all existing quantum field theories that have empirical corroboration are presumably long-range (and therefore high-dimensional-representation) approximations to an exact theory of interacting qubits. Note that the solution given in Section 3, having a 4-dimensional representation, presumably describes a qubit field only at very short ranges, or perhaps at very low temperature when very few of its degrees of freedom are excited, and therefore not in a regime in which any such approximation holds.

Such discussion can only be speculative at present. Of more immediate importance is that it is not obvious that the theories described in this paper satisfy the original motivation of the investigation: that the information-carrying capacity of a system of finite volume should be finite. For although in these theories such systems have finite-dimensional Hilbert spaces, I have not proved that the connection between information-carrying capacity and Hilbert space dimension is the same in these theories as it is in the conventional theory. I conjecture that it is, but proving that is beyond the scope of this paper: it requires an extension of the theory to interacting qubit fields, and then to a theory of measurement and a theory of computation for qubit fields, both of which must be significantly different from their existing counterparts. Only when those theories have been developed can we investigate what information is, and how it flows, in this new type of quantum field theory.